\UseRawInputEncoding

\documentclass[12pt]{article}
\usepackage{times}

\usepackage{amssymb, amsmath,float}
\usepackage{natbib}
\usepackage{graphicx,color}
\usepackage{subfig,microtype}
\usepackage{lineno}

\usepackage{setspace} 
\usepackage[breaklinks]{hyperref}

\usepackage{memhfixc}

\usepackage[labelfont=bf,labelsep=period,justification=raggedright]{caption}

\makeatletter
\renewcommand{\@biblabel}[1]{\quad#1.}
\makeatother

\date{}

\title{\bf{Social cohesion V.S. task cohesion: An evolutionary game theory study
} }

\author{  Xinglong Qu$^{1,^\star}$ , {Shun Kurokawa}$^{2}$, The Anh Han$^{3}$} 

\date{\normalsize \today}
\begin{document}

\maketitle
{\footnotesize
\noindent
$^{1}$ The Research Center of Information Technology \& Social and Economic Development, Hangzhou Dianzi University, Hangzhou, Zhejiang, China\\
$^{2}$ Graduate School of Arts and Sciences, University of Tokyo, 3-8-1 Komaba Meguro-ku, Tokyo 153-8902, Japan \\
$^{3}$ School of Computing, Engineering and Digital Technologies,  Teesside University, Borough Road, Middlesbrough, UK TS1 3BA  \\
$^\star$ Corresponding author: Xinglong Qu (quxinglong@amss.ac.cn)
}


\section*{Abstract} 
Using methods from evolutionary game theory, this paper investigates  the difference between social cohesion and task cohesion in promoting the evolution of cooperation {in group interactions.
{Players engage in public goods games} and are allowed to leave their groups if too many defections occur.
Both social cohesion and task cohesion may prevent players from leaving.
{While a higher level of social cohesion increases a player's tolerance  towards defections,  task cohesion is associated with her group performance in the past.}
{With a higher level of task cohesion, it is more likely that a dissatisfied player will refer to the history and  remains in her group if she was satisfied in the past}.
Our results {reveal} that social cohesion is {detrimental to the evolution of} cooperation while task cohesion facilitates it.
This is because social cohesion hinders the conditional dissociation mechanism but task cohesion improves the robustness of cooperative groups which are usually vulnerable to mistakes.
We also discuss other potential aspects of cohesion and how {they can be investigated through our modelling}.
{Overall, our analysis provides novel insights into} the relationship between group cohesion and group performance through studying the group dynamics and suggests further application of evolutionary game theory in this area.

\textbf{Keywords:} public goods game, cooperation, evolutionary game theory, conditional dissociation, social cohesion, task cohesion


\section{Introduction} 
\label{text:intro}

{Group cohesion or cohesiveness is one of the oldest and most widely studied factors in the group dynamics}
{literature \citep{Campbell2009Sticking,Mullen1994The}.}
Despite a {large body of works  showing} that group cohesion is positively related to the group performances, such as  job satisfaction, psychological well-being, and group efficiency \citep{Beal2003Cohesion,Carless2000The,Mullen1994The}, there are also some controversies \citep{Friedkin2004Social,Dyaram2005Unearthed,Esa2009can,Khoshsoroor2019Team}. That includes groupthink \citep{Janis1982Groupthink}, a conjecture that the effort to  {reach} an agreement may lead to  {a} dysfunctional decision-making outcome.
This inconsistency highlights the fact that our understanding of the mechanism of how group cohesion works on the group performance is insufficient \citep{Carron2000Cohesion,Campbell2009Sticking,Mcleod2013towards}. Notably, \citet{Drescher2012Cohesion} even call it a spectacular embarrassment in group theory and research.
While {apparently} the complex nature of cohesion is a major cause of such conflicting results, there exist other crucial drawbacks in the current literature.

The {principal} one rests in the inconsistency of the definition and measurement of cohesion among researchers.
To make it up, \citet{Friedkin2004Social} and \citet{Campbell2009Sticking} suggest scientists to turn focus on what is concrete and its interrelationship with other constructs.
One milestone is the distinction between social and task cohesion \citep{Mikalachki1969Group,Dion1992Cohesiveness,Campbell2009Sticking}.
The social facet includes relationships within the group,  {while} the task facet includes collective performance, goals, and objectives \citep{Carron1985The,Carron2000Cohesion}.
In general, it is found that task cohesion is more positively related to group performance than social cohesion \citep{Zaccaro1991Nonequivalent,Bernthal1993Cohesiveness,Mullen1994The}.
But unfortunately, the explanation for it is in short either.

Another shortage in the conventional studies is the overlook of the dynamic nature of groups \citep{Mathieu2008Team,Campbell2009Sticking,Drescher2012Cohesion, Mathieu2015Modeling}. 
In reality, individuals continually join and leave different groups and in the meanwhile interact with each other. 
However, most of the previous works fail to describe this dynamical process, {and more importantly,} its influence on peoples' decision-making.
\citet{Mathieu2015Modeling} contribute  {one of the most important works which proves} positive reciprocal relationships between cohesion and performance over time. 
They firstly synthesise aforetime studies that are about the reciprocal influence of group cohesion and performance. 
Then they apply  {the so-called} structural equation (SEM) method to analyse the results, and justify their model with data collected from a series of business simulations that last over 10 weeks.
While SEM is inspiring, scholars are cautious about drawing strong causal inference from it \citep{Mathieu2015Modeling}, and serious scrutiny is required before assuming the framework of the model\footnote{  \citet{Mathieu2015Modeling} find only 17 papers {that} use SEM to study the reciprocal effect of cohesion and performance, which is the most key question in the study of cohesion, not to  {mention} other aspect of cohesion.}.

To further reveal the  mechanism regarding how cohesion impacts group performance, we take players' decision-making into consideration using an evolutionary game theory model.
In the later half of last century, game theory was developed as a fundamental tool to analyze the conflicts and cooperation among human society and other organisms \citep{Aumann2019Lectures}.  
However, there are multiple deficiencies lie in the traditional theory of games that centres on the calculation and analysis of equilibrium, such as complexity of the calculation, equilibrium selection, hyperrational agents and so on.
These problems motivate social scientists in a variety of areas to turn their eyes on evolutionary game theory \citep{Alexander2019Evolutionary}.
It assumes the decision makers are short sighted, their action patterns are heuristic, and very limited information is available.
Despite the simple rules and assumptions, its effectiveness in explaining the world wasn't overshadowed.
Especially when considering large population behavior, simple action rules could generate complex social phenomena which coincide with the institution and social facts \citep{Newton2018Evolutionary}.

As aforementioned, while the complex nature of cohesion makes it hard to draw an overall conclusion, it is meaningful to focus on the concrete aspects of cohesion and apply different theoretical tools to study it. 
{An important characteristic of cohesion could be drawn from the definition of \citet{Festinger1950Informal} that it is "the resultant forces which are acting on the members to stay in a group€.
That is, the more cohesive a group is, the more likely that its members would stay within it.
However, people in reality keep leaving and joining new groups for different reasons.
In game theory study, this phenomenon is studied in the model of conditional dissociation, which allows dissatisfied players to break off with their opponents.
This mechanism is shown to promote cooperation in the two-person prisoners' dilemma game \citep{Aktipis2004,Izquierdo2010,Qu2016Conditional}.
\citet{Qu2019How} extend this model to the public goods game, which is a multi-person game, and also incorporate cohesion into consideration.
In the model, if cohesion exists, group members who do not want to leave would be united together, otherwise, all group members become single.
Their main finding is that cooperation is better promoted when cohesion exists as it results in  {a greater  chance} for cooperators to play with each other, which {is} usually termed as positive assortativity \citep{Eshel1331,doebeli2005models,fletcher2006altruism,Fletcher2009A}.}

In this paper, we {examine other important} aspects of cohesion in influencing group cooperation. 
In particular, as  {motivated above and also emphasized by} \citet{Mathieu2015Modeling}, we compare social and task cohesion.
Our basic model is  in line with \citet{Qu2019How} where a well mixed  {population of players interacting with each other using the one-shot}  public goods game under conditional dissociation  {mechanisms}. 
Players' leaving decisions depend on both their own tolerance towards defections and the levels and types of cohesion.
Social and task cohesion are inspected separately in the first two settings and synthetically  {afterwards}.
When some players are unsatisfied, both types of cohesion may prevent the group from dismissing. 
{More concretely}, with social cohesion, players tend to become more tolerant toward defections; while with task cohesion,
dissatisfied players become more patient only if  {their goal was achieved in the past}.
To be more specific, while social cohesion is described as forces irrelevant to the history of interactions, task cohesion consists in the likelihood that a player might refer to the play history in  {the} last round.

Our model reveals that, social cohesion and task cohesion, defined as above,  {have} opposite effects on the evolution of cooperation  {where} the former one harms it but the latter benefits it.
The main reason accounting for the failure of social cohesion and success of task cohesion in our model is that social cohesion counteracts while the latter enhances the positive assortativity effect \citep{Qu2019How}.
With either type of cohesion, the dissociation mechanism is hindered and players would stay together longer.
So the higher the levels of cohesion are, the less likely that dissociation happens, which explains the negative effect of social cohesion.
However, with task cohesion, players would leave their groups for sure once they are dissatisfied in two continuously rounds.
Moreover, since the winning cooperative strategy is intolerant towards defection, it is vulnerable to mistakes. 
Task cohesion prevents groups from  {being dismissed by mistakenly defecting}, which keeps the cooperative  {groups} play together longer.

This finding partially in line with the vast empirical literature that claim task cohesion is more positively related to group performances than social cohesion \citep{Zaccaro1988Cohesiveness,Zaccaro1991Nonequivalent,Mullen1994The,Chang2001A, Warner2012Team,Spink2014Group}.
Our finding supports the common wisdom  that organizations and practitioners should encourage team members to share successful experience more often.
Intuitively, this would increase the attraction towards teams or group pride or morale, which according to our analysis, also  {increases} the task cohesion that bonds group members based on shared commitment to achieving goals.

Despite the difference of our results from the classical conclusion about the  positive significance of social cohesion, we  {will highlight} other potential aspects of cohesion that  {may be} associated with group performance.
When players are less likely to update their strategies, cooperation is higher.
In reality, people update their strategies to improve their utilities.
High levels of cohesion is usually associated with higher levels of job satisfaction, which thus reduces the possibility they update strategies.
As the probability of making mistakes decreases, group cooperation increases.
In reality, it is reasonable to conjecture that groups with higher levels of social cohesion would communicate more effectively and accurately and thus make less mistakes, which improves group performances.
Another more subtle one may be that social cohesion is related with selection intensity.
In evolutionary game theory study, selection intensity is an important element of the evolution dynamics.
\citet{Rand2013Evolution} find that a moderate level of selection intensity is optimal for the emergence of generosity behavior.
In their experiment, they come up with a sagacious measurements for selection intensity by asking subjects that among those they interact with in daily life, how clear is it which people are more or less successful?.
The more clear their answers are, the larger the selection intensity is implied.
So cohesion  {might be} important in maintaining a relatively optimal level of selection intensity.

The rest of this paper is organized as follows. We detail our model in Section \ref{section-model} and present the simulation results in Section \ref{section-results}. We then briefly examine and discuss  other potential aspects of cohesion in Section \ref{section-theoretical}. The conclusions are described in Section \ref{section-conclusion}, with some further discussion being followed  in Section \ref{further}.

\section{Models}
\label{section-model}
\subsection{Public goods game}
Consider a finite population of individuals who are going to play the following public goods games.
$G$ persons are grouped together with everyone being endowed with 1 unit of personal token.  
They simultaneously decide whether or not to contribute their personal token to the public wealth which would magnify by $r(1<r<G)$ times and be equally shared among all the group members. 
As the reserved personal token would not increase, the payoff function for a player is defined as:
\begin{equation}
	u_i=\frac{r}{G} \cdot \sum\limits_{j=1}^{G} a_j+(1-a_i),
	\label{eqpayofffunction}
\end{equation}
where $a_j=1$ if player $j$ cooperates by contributing and $a_j=0$ if she reserves her personal wealth in which case we say she defects.

\subsection{Conditional dissociation and strategies}
Conditional dissociation mechanism \citep{Izquierdo2010,Qu2016Conditional,Qu2019How} has proven to be an efficient mechanism to promote the evolution of cooperation.
Initially, all players are single players and they are randomly grouped to play the games.
After each round of game, if a player feels unsatisfied about her group, under conditional mechanism, she could decide whether or not to stay with her current group opponents.
By leaving, both she and her group mates become single and enter the matching pool. 
All the single players in the pool would be randomly regrouped to play the games in next round of game.

So, for each player $i$, her strategy is denoted by a tuple $(a_i,b_i)$ where $a_i\in \{0,1\}$ indicates whether she is a cooperator $(a_1=1)$ or defector $(a_1=0)$ when playing the game, and $b_i\in \{0,1,\cdots, G-1\}$ indicates her tolerance towards defections.
Denote $\Sigma=\{0,1\}\times\{0,1,\cdots, G-1\}$ as the strategy space for agents.
We assume mistakes may happen to each agent randomly with a fixed probability of $\epsilon$, that is, a player who intended to cooperate may wrongly defect and a defector may wrongly cooperate.
Let $d_i$ denote the number of defectors observed by player $i$. 
Players are satisfied with their groups and choose to stay within current group only if 
\begin{equation}
	d_i\leq b_i.
	\label{originalcd}
\end{equation}
A group dissolves if any of its members are dissatisfied and choose to leave.
After a group dissolves, all its members, whether satisfied or not in the last round, become single players and enter the matching pool.
Players in the matching pool regroup with each other randomly and play in the next round.

In the above model, whenever some players are dissatisfied, their groups dissolve immediately.
In this paper, different types of cohesion that prevent the dissolution of groups are investigated.

\subsection{Social cohesion and task cohesion}

We compare two constructs of cohesion, i.e. \textit{social cohesion} and \textit{task cohesion}.

{Social cohesion} is the attractiveness that players feel about their groups, which is independent of the history of games.
To be more specific, for each player $i$, her perception of the {social cohesion}, denoted by $\xi_i=B(G,p)$, is a random variable satisfying a Binomial distribution $p$ indicates the level of social cohesion. 
A player is satisfied if the number of defections is no more than the sum of her tolerance and her perceived social cohesiveness, that is,
\begin{equation}
	d_i\leq \xi_i+b_i.
	\label{eqsatisfaction}
\end{equation}
Compared with \eqref{originalcd}, social cohesion $\xi_i$ is added to the right side. Apparently, the expected cohesion $E\xi_i=G\cdot p$ so the higher level of cohesion the more likely that player stays within the group.
Here we assume $\xi_i$ to be random since it is player's perception or feelings which usually are affected by some other external factors and  change over time.
After each round of play, if a player is dissatisfied in the sense that she observes too many defections than her tolerance and social cohesion, she chooses to leave and all her group members enter the matching pool.

Task cohesion is dependent on how well players achieve their goals while playing together.
If a group of players have played together in the last rounds, players may refer to the history of the last round of game.
And the level of task cohesion is defined here as the probability that players refer to the history.
To be specific, with task cohesion being $q$, a dissatisfied player (i.e. when the condition in Eq 2 is violated) would choose to leave the current group with probability $1-q$ and would look back on the last round with probability $q$.
If she was satisfied in the last round, then she would still choose to stay within current group. 
The higher the task cohesion is, the more likely that player would refer to the history before they make their final decisions about leave or stay.
When $q=0$, players do not look back on the history and leave their current group immediately if they are dissatisfied.
When $q=1$, players always refer to the history and stay within the current group unless they are dissatisfied in two consecutive rounds.

In our work, we investigate the role of social cohesion and task cohesion separately in the first two settings and then jointly in the third.
\subsection{Evolution dynamics}

After each round of game, every player updates her strategy with a probability of $\delta$.
If a player updates her strategy, she would learn the strategy from another player in the population.
The learning process is a Moran process \citep{Nowar2004Emergence,Taylor2004Evolutionary,Szabo2007evolutionary}, which means the probability that one player's strategy is learned by others is  $\frac{f_i}{\sum f_i}$ where $f_i$ is player's fitness defined as:
\begin{equation*}
	f_i=u_i^s.
\end{equation*}
Here $u_i$ is the player's payoff\footnote{Payoffs gained in each round are consumed immediately and cannot be accumulated.} and $s$ measures the selection intensity: when $s=0$, the process is just the neutral drift and when $s\rightarrow\infty$, it is the best-response dynamics.
Meanwhile, mutations may also happen with a fixed probability $\mu$ to the learner that she would randomly choose a strategy from the strategy set $\Sigma$ regardless of its performance.
The strategy update in each round is simultaneous which means if a player has just updated her strategy and is learned by others, it is her former strategy rather than the new strategy could be learned.

\section{Results}
\label{section-results}
Our main results are derived from computer simulations.
We run three series of simulations to compare how social cohesion and task cohesion affect the group cooperation.

\begin{figure}[!htb]
	\centering
	\subfloat[Evolution of cooperation]{{\includegraphics[width=1\linewidth]{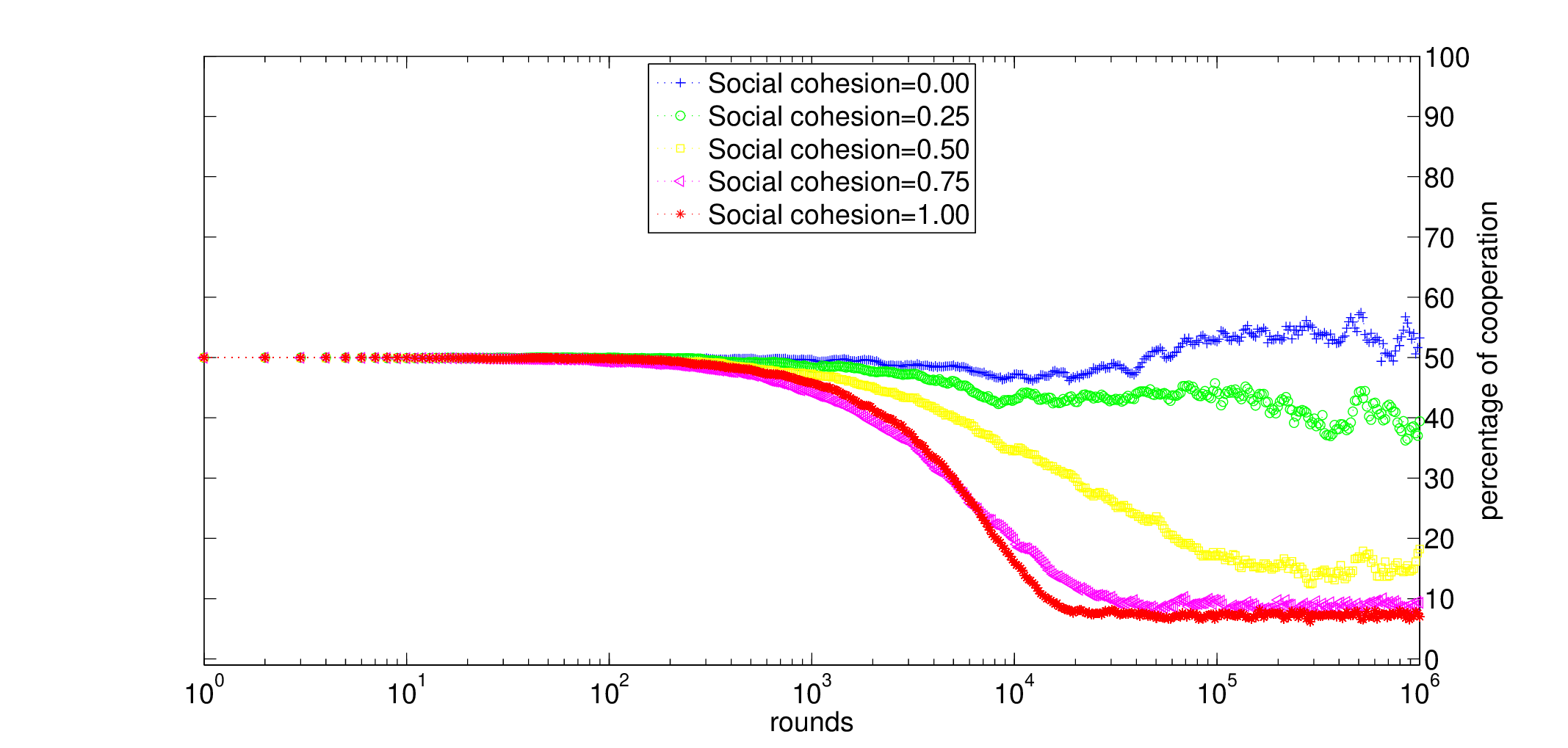}}}
	\qquad
	\subfloat[Distribution of strategies]{{\includegraphics[width=1\linewidth]{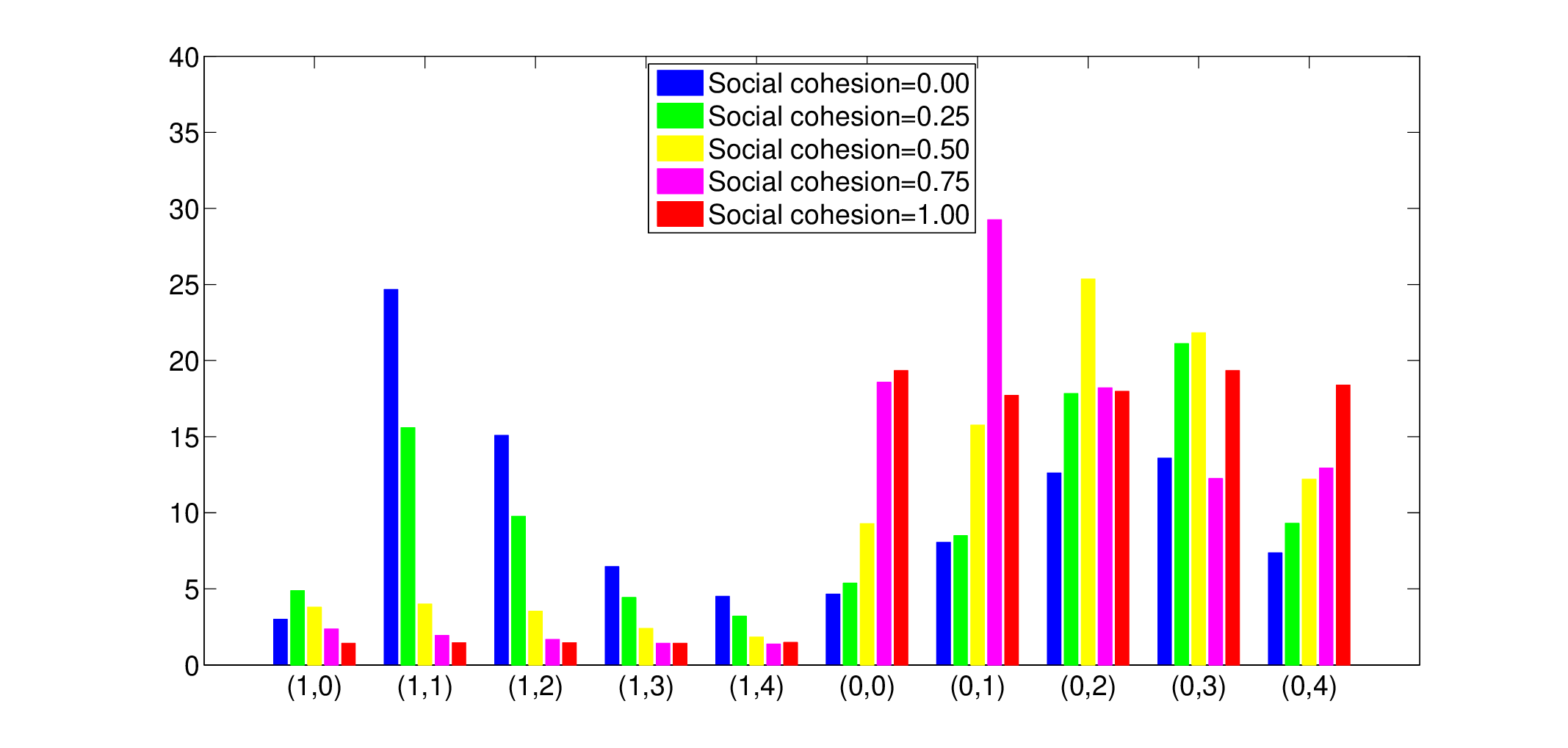}}}
	\caption{The evolution of the percentages of cooperation after $10^6$ rounds of games for different levels of social cohesion. Mistake=0.01, Group size =5, population $N=200$, mutation rate $\mu$=0.05, strategy updating rate $\delta=0.001$, r=3, selection intensity $s=1$.}
	\label{fig:socialcohesion}
\end{figure}
\subsection{Only social cohesion exists}
Firstly, we investigate how different levels of social cohesion influence the emergence of cooperation.
In this set-up, players' decisions only depend on the current status of play.
They leave if their tolerance and the attractiveness of their group are unable to make them satisfied.

As can be seen from Figure \ref{fig:socialcohesion}(a), the stronger social cohesion is, the less cooperation is observed.
When social cohesion is fully hypothesized, only about 10 percent of players are cooperators.
This verifies that conditional dissociation is beneficial  to the evolution of cooperation from another direction but contradicts the general hypothesis that social cohesion is good for group performances.
As when social cohesion increases, conditional dissociation mechanism is weakened and thus cooperation declines.

We can see from Figure \ref{fig:socialcohesion}(b) that when cooperation flourishes, the most successful strategy is $(1,1)$ and the least tolerant strategy always performs badly.
However, as strategies become more tolerant, they perform worse. 
The strategy distribution is quite different to those observed in \citep{Qu2019How}, where the only winning cooperative strategy is the most intolerant one i.e.$(1,0)$.
One major difference is that in their model, whenever some players update their strategies, their groups dismiss.
Compared with $(1,0)$ strategists, players using $(1,1)$ settle down more quickly as they could tolerate one more defector.
At this point, despite being exploited by the single defector, their payoffs are higher than the average of the whole population and they are more likely learned by other agents, which means they are more successful in the evolution.

\subsection{Only task cohesion exists}
\begin{figure}[!htb]
	\centering
	\subfloat[Evolution of cooperation for different task cohesion]{{\includegraphics[width=1\linewidth]{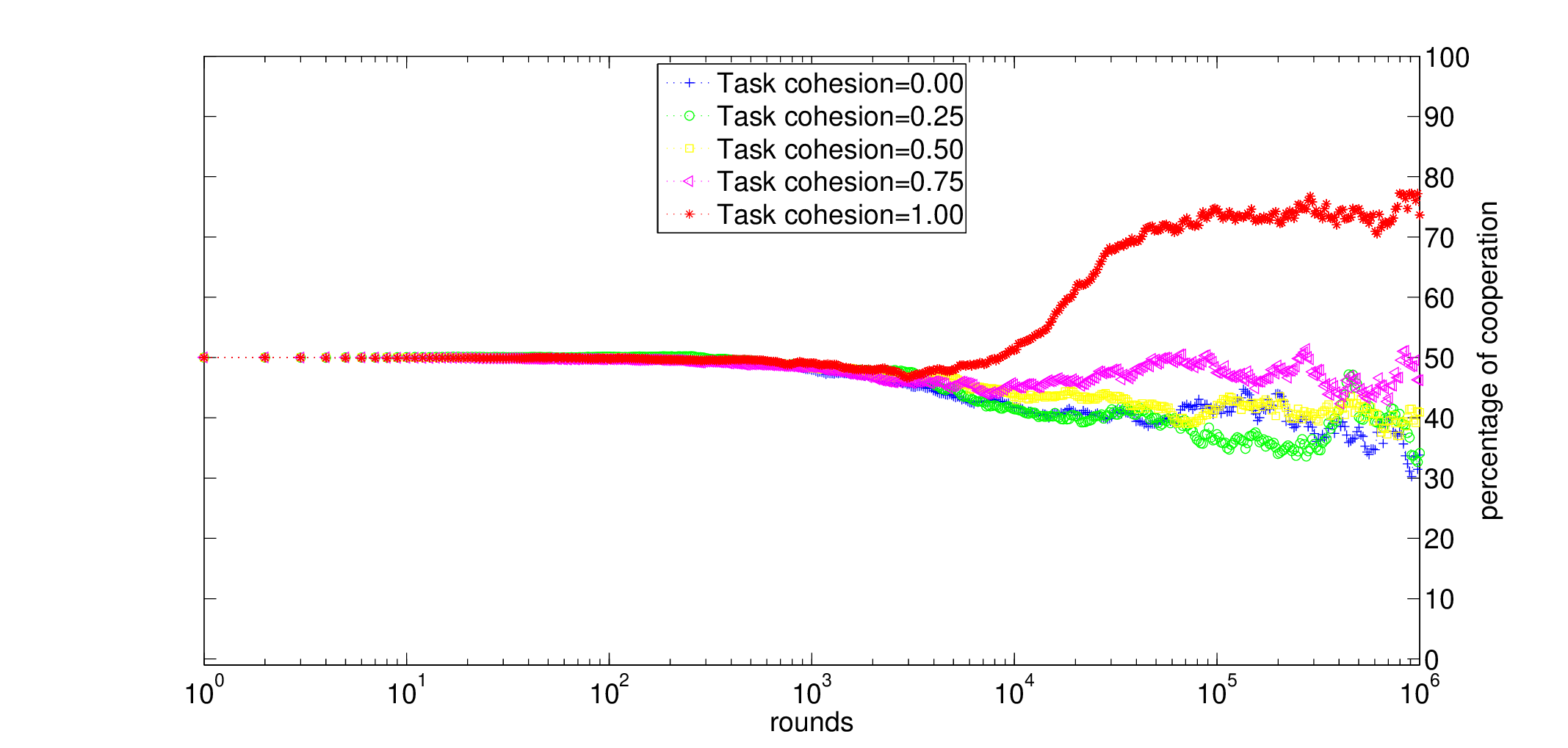}}}
	\qquad
	\subfloat[Evolution of cooperation for different  task cohesion]{{\includegraphics[width=1\linewidth]{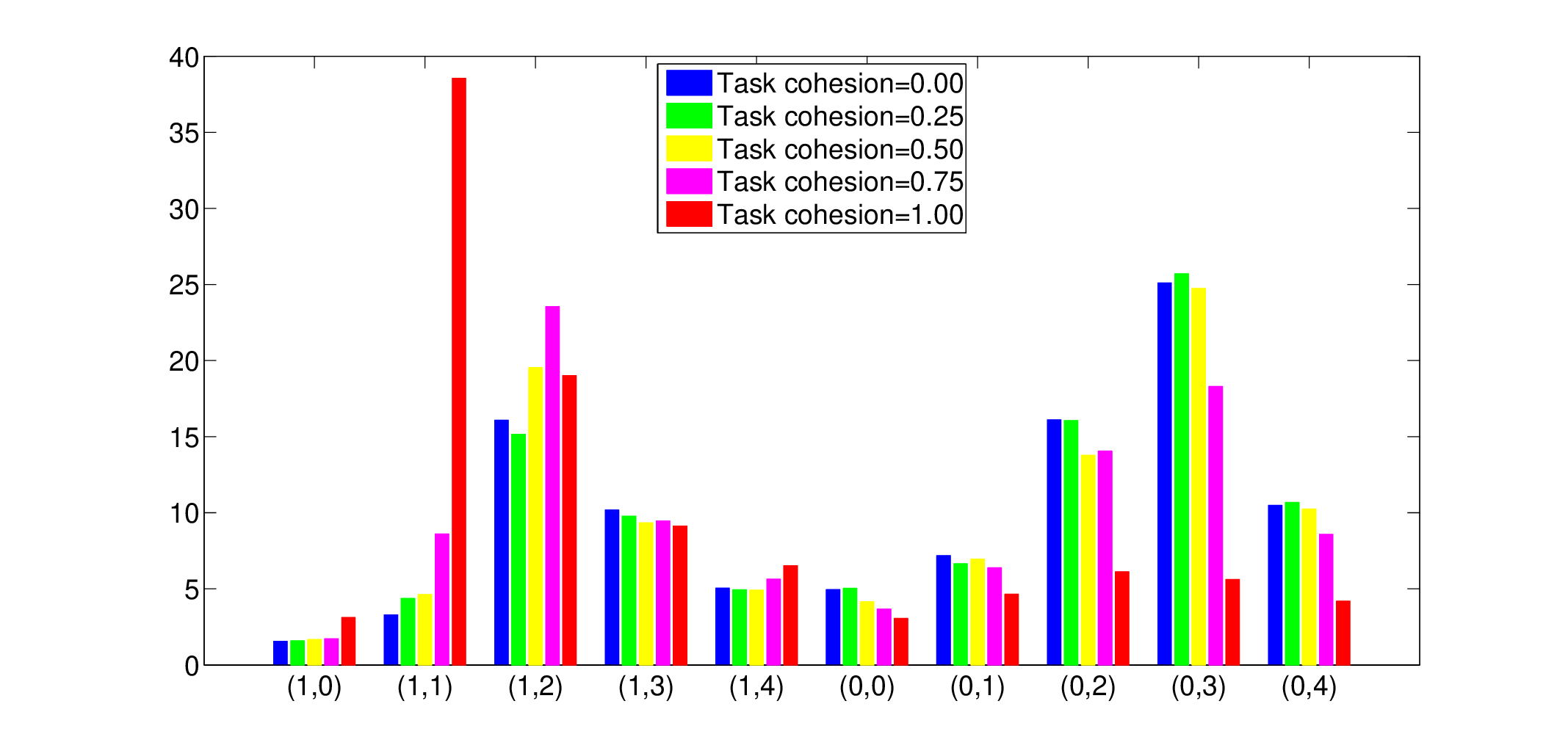}}}
	\caption{The evolution of the percentages of cooperation after $10^6$ rounds of games for different levels of task cohesion. Mistake=0.05, Group size =5, population $N=200$, error rate=0.05, strategy updating rate $\delta=0.001$, r=3, selection intensity $s=1$.}
	\label{fig-taskcohesion}
\end{figure}
With task cohesion, individuals may make their leaving decisions based on both the current and the previous rounds of play.
If they are satisfied in either round, they would stay.

As can be seen from Figure \ref{fig-taskcohesion}, the effect of task cohesion is 
indistinguishable when cohesion is smaller than 0.5. However, as task cohesion increases, cooperation increases substantially.
The cooperation level is highest when players always consider the play history in the last round.

The strategy distribution is similar to the situation where only social cohesion exists.
When task cohesion is 1, the winning cooperative strategy is also $(1,1)$.
A tiny difference is that, except for $(1,1)$ that performs better and better as task cohesion increases, all the remaining cooperative strategies  perform almost the same in different settings.

Since positive assortativity is the explanation for how dissociation mechanism promotes cooperation, it would be natural to investigate how task cohesion enhances it.
In the study of conditional dissociation mechanisms, players are rather impulsive that whenever they feel dissatisfied, they would leave their opponents immediately.
Intuitively, when a group of cooperators meet together, they would continue playing together until someone defects by mistake.
In our model, with higher levels of task cohesion, players get more discreet before they decide whether to stay within the current group or not.
And a pleasant history of play would stop the dissolution of groups.
So with higher levels of task cohesion, cooperative groups are less likely to dissolve which means enhanced positive assortativity.
\begin{figure}
	\centering
	\includegraphics[width=1\linewidth]{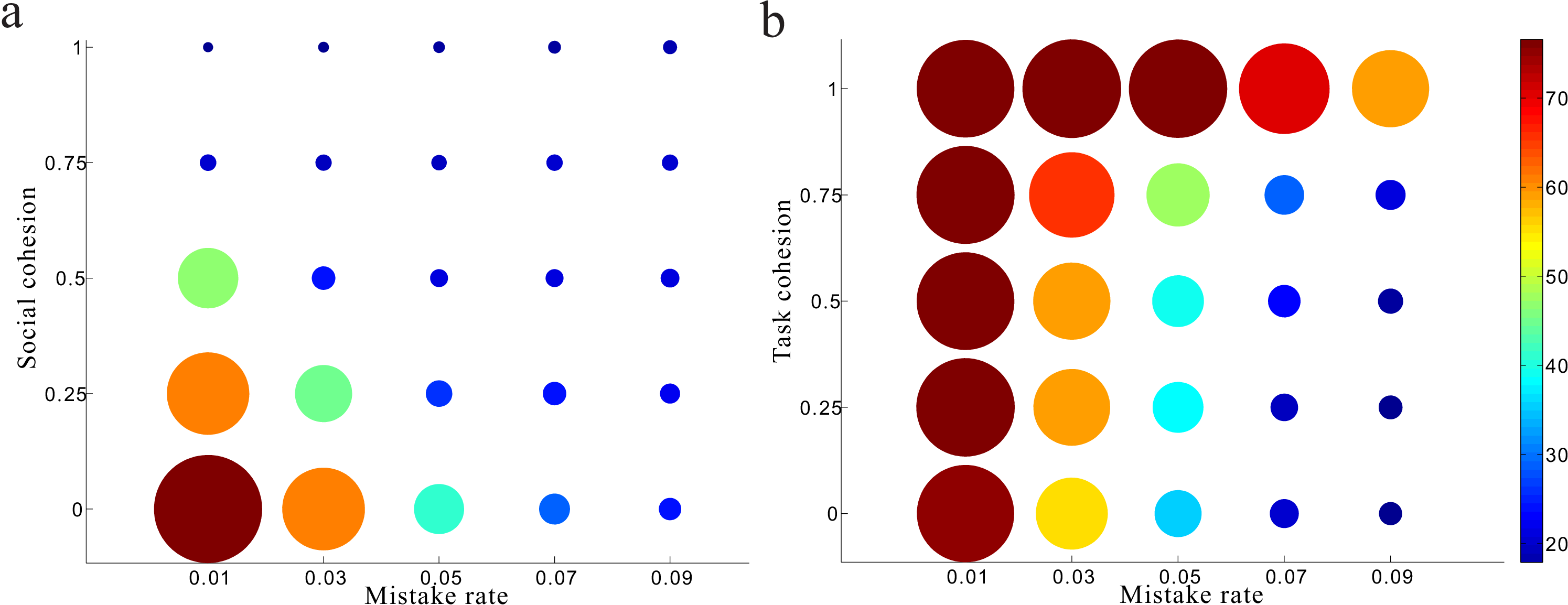}
	\caption{The percentage of cooperation after $10^6$ rounds of simulations with different levels social cohesion, task cohesion and mistake rate.
		The colors and sizes represent the percentages of cooperation. The more cooperation, the larger the circle filled with the hotter color. Group size =5, population $N=200$, mutation rate $\mu$=0.05, strategy updating rate $\delta=0.001$, r=3, selection intensity $s=1$. }
	\label{fig:contour-mist-task2}
\end{figure}
{In Figure \ref{fig:contour-mist-task2}, we present the results of evolution for different levels of social cohesion, task cohesion and mistake rates.
It can be easily seen that cooperation rates always decrease as mistake rates increase.
Both types of cohesion reduce the differences of cooperation rates under low mistake rates and high mistake rates but at different directions.
When the mistake rate is low, the effect of increasing  task cohesion is rather limited while it is obvious that increasing social cohesion is detrimental to cooperation. 
When the mistake rate is higher, the advantage of higher level task cohesion becomes more obvious, which suggests that task cohesion is effective in preserving cooperation from mistakes.}

\subsection{Both social and task cohesion exist}
We now combine both types of cohesion and examine whether there is any synergy effect from the combination.
In each round of the game, player $i$ is satisfied if and only if the number of observed defections $d_i$, her tolerance, and the perceived social cohesiveness satisfy equation \eqref{eqsatisfaction}.
With task cohesion, every player might also refer to the history to determine whether or not to leave.
If a dissatisfied player doesn't refer to the history, or she was also dissatisfied in the last round, she leaves.
Otherwise, a player would remain in her group if she is satisfied in current round or she refers to the history and finds herself satisfied in the last round.

Our results suggest the negative effect of social cohesion and positive effect of task cohesion remain when both types of cohesion exist.
In Figure \ref{fig:contourplot}, we present the outcomes for different levels of social and task cohesion. 
For any levels of task cohesion, the proportions of cooperation decrease with increasing social cohesion, which indicates the negative effect of social cohesion.
On the other hand, if we increase the levels of task cohesion, we observe higher levels of cooperation with higher levels of task cohesion.
So task cohesion promotes cooperation whatever levels of social cohesion are.

The different {effects for the evolution of cooperation in group interactions from} the two types of cohesion further support the idea to {differentiate  between them in real-world social and economic contexts}.
\begin{figure}[tbh!]
	\centering
	\includegraphics[width=1\linewidth]{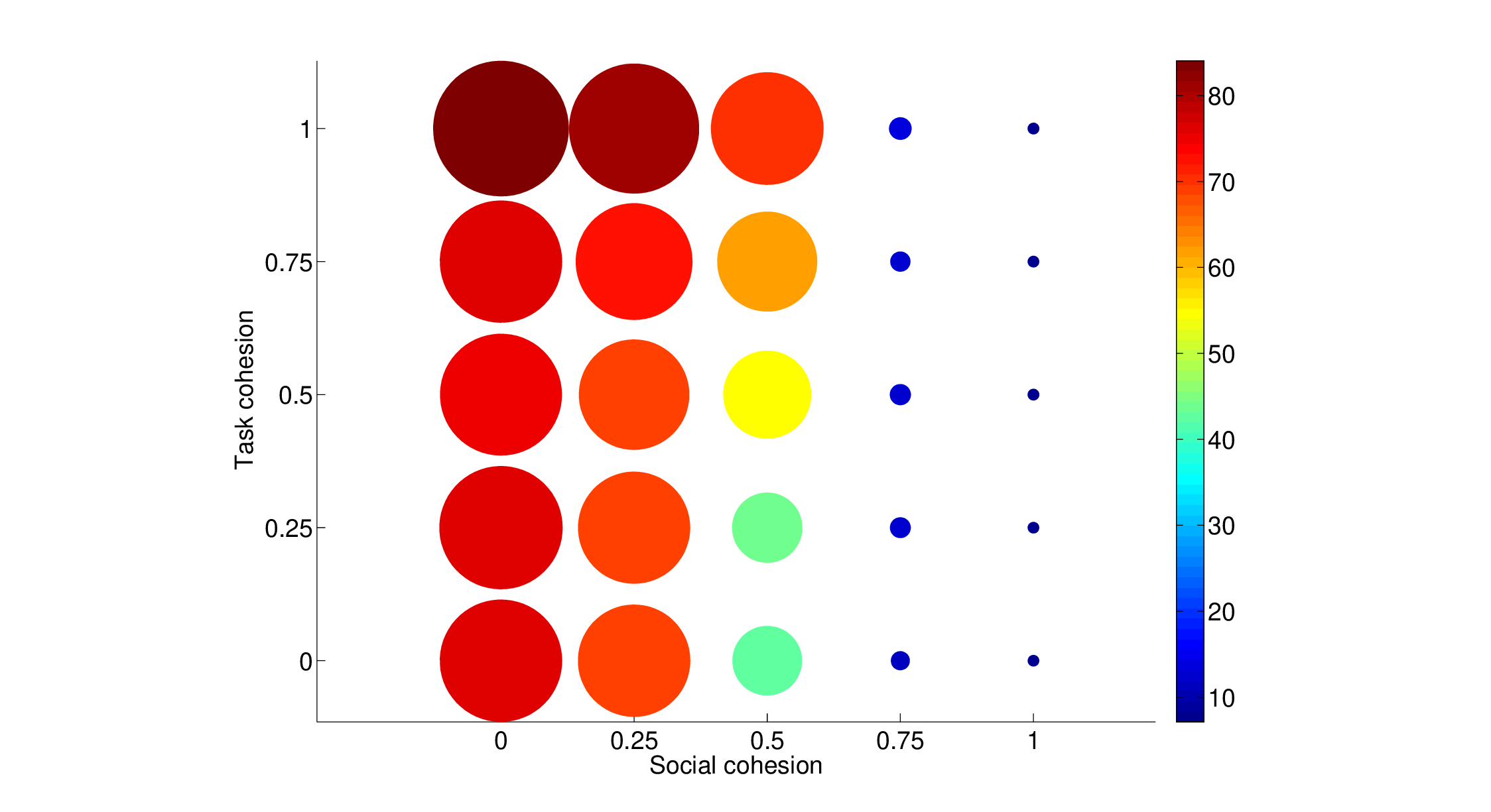}
	\caption{The percentages of cooperation after $10^6$ rounds of games for different levels of social and task cohesion. Mistake rate=0.01, Group size =5, population $N=200$, mutation rate $\mu$=0.05, strategy updating rate $\delta=0.001$, r=3, selection intensity $s=1$.}
	\label{fig:contourplot}	
\end{figure}

\clearpage
\section{Potential aspects of cohesion}
\label{section-theoretical}
While contradicting the common viewpoint that suggests positive relationship between 
social cohesion and group cooperation, our model sheds light on other potential aspects of cohesion, which could improve group performances but are not involved in the current model.
\subsection{Strategy updating and mistake rate}
Our model confirms with the classical conclusion that cooperation is more abundant under lower strategy updating rate (Figure \ref{fig:contour-mist-task2}) and mistake rate (Figure \ref{fig:group5-mu0}).

Cohesion may lead to lower levels of strategy updating rate since it increases individuals' sanctification.
In human society, people change their strategies continually for different reasons.
The primary one would be to pursue better outcomes so they learn it from more successful ones.
With higher levels of group cohesion, players are  more easily to be satisfied, so their incentives to pursue better outcomes decline, and update their strategies less.
\begin{figure}
	\centering
	\includegraphics[width=1\linewidth]{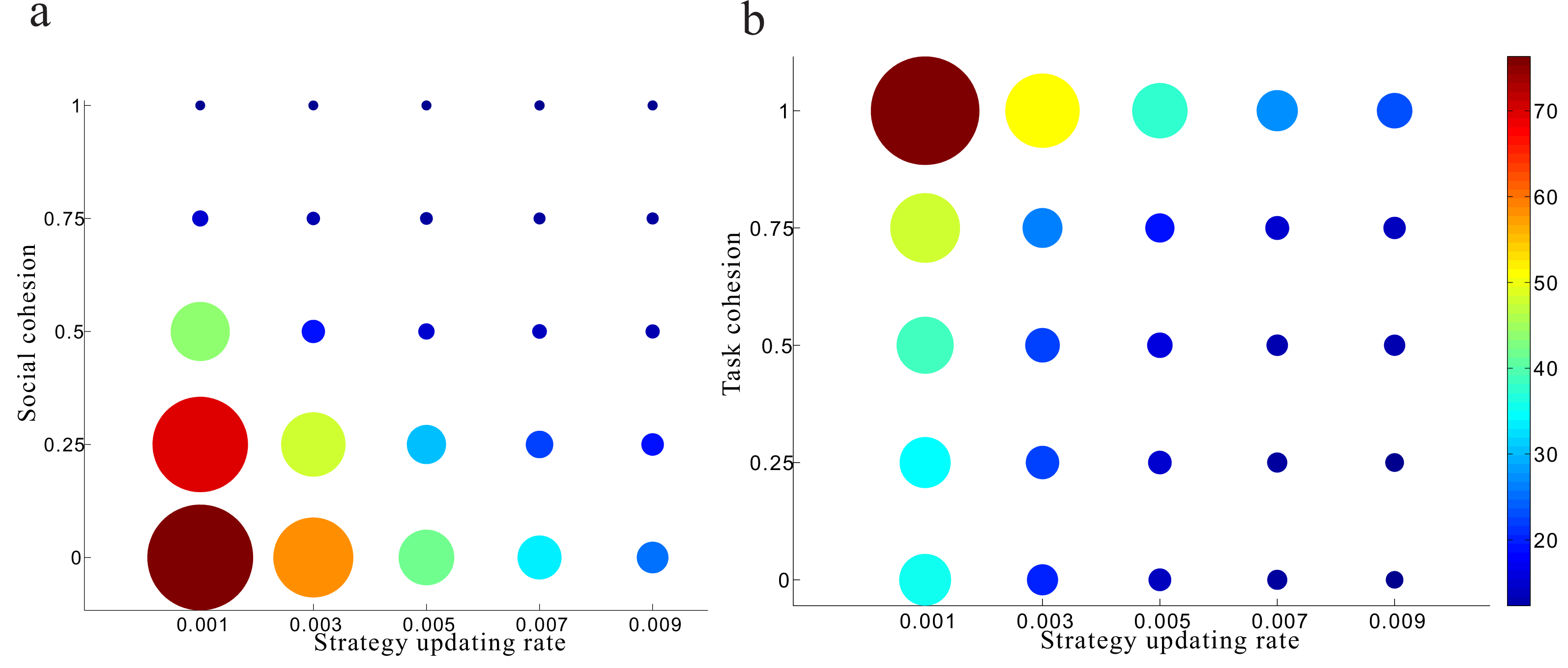}
	\caption{The percentage of cooperation after $10^6$ rounds of simulations with different levels social cohesion, task cohesion and strategy updating rate.
		The colors and sizes represent the percentages of cooperation. The more cooperation, the larger the circle filled with the hotter color. Group size =5, population $N=200$, mutation rate $\mu$=0.05, selection intensity $s=1$, r=3. The mistake rate is $0.01$ in figure a for social cohesion and  it is 0.05 in figure b for task cohesion.}
	\label{fig:group5-mu0}
\end{figure}

In reality, everyone makes mistakes but by being more careful, they are less likely to do so.
The sense of responsibility has been emphasized by some other studies \citep{Chan2006reconsidering,Dickes2010construct,Schiefer2017The} on cohesion.
With higher cohesion, players feel more responsible to act carefully, which thus could  reduce the probability that individuals make mistakes.
As the winning cooperative strategy is also intolerant of defections, less mistakes help them to play together longer and gain more benefits. 

\subsection{Selection intensity}
In the study of evolutionary game theory, selection intensity has determinant influence on the outcomes.
However, in reality, how it is related to human society or even natural world is not so clear.
\citet{Rand2013Evolution} measure selection intensity by asking subjects how clear to tell that someone is more successful.
This measurement obviously is related to the social and cultural aspects of our society, which have non-trivial influences on human behavior.
Inspired by \citet{Rand2013Evolution}, our model also suggests that social cohesion may forge a medium level of selection intensity is better for group cooperation.
\begin{figure}
	\centering
	\includegraphics[width=1\linewidth]{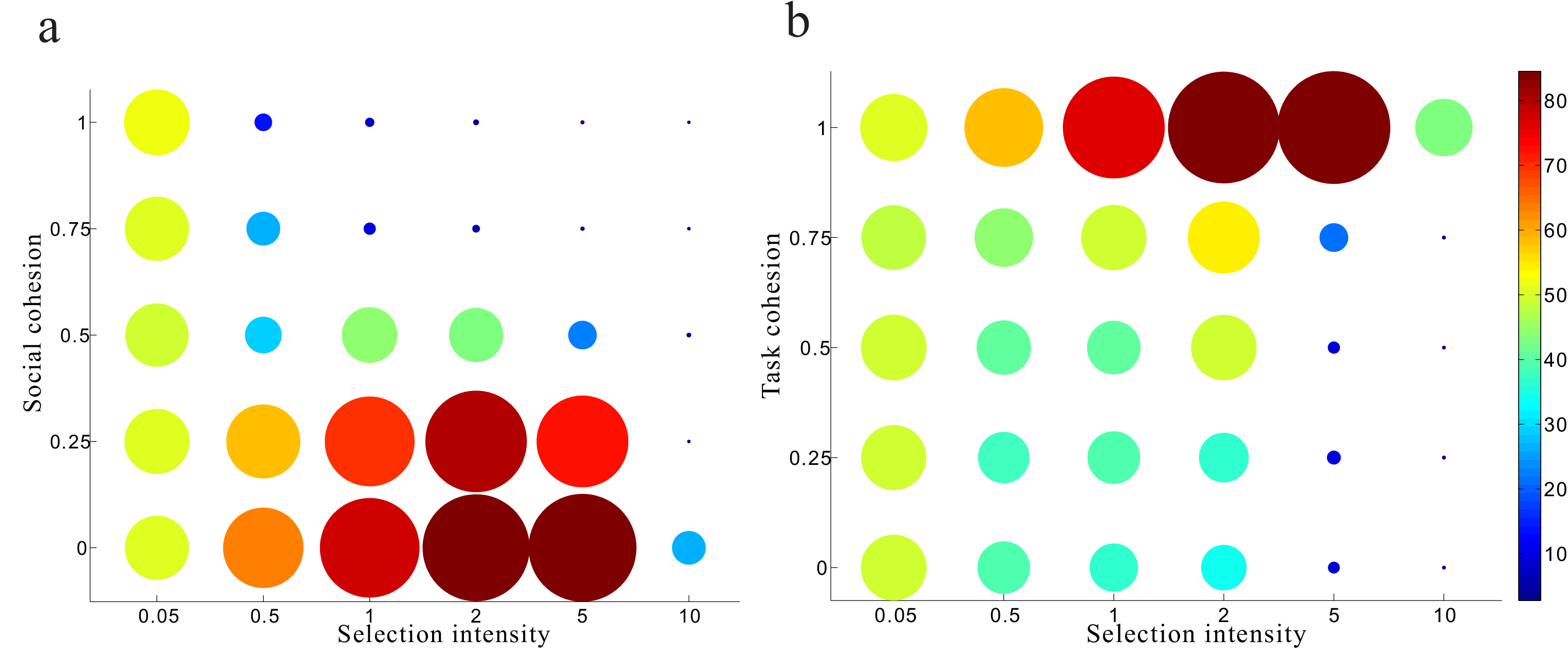}
	\caption{The percentage of cooperation after $10^6$ rounds of simulations with different levels social cohesion, task cohesion and selection intensity.
		The colors and sizes represent the percentages of cooperation. The more cooperation, the larger the circle filled with the hotter color. Group size =5, population $N=200$, mutation rate $\mu$=0.05, strategy updating rate $\delta=0.001$, r=3. The mistake rate is $0.01$ in figure a for social cohesion and  it is 0.05 in figure b for task cohesion.}
	\label{fig:sltstrth}
\end{figure}

{It is worthwhile to remind that, with tiny selection intensity (say it is being 0.05), the distributions of strategies  remain {almost}  unchanged, so we see similar levels of cooperation (near 50\%) in {all} settings in Figure \ref{fig:sltstrth}.
	When only {a low level} of social cohesion exists, cooperation rates firstly increase and then decrease as the selection intensities increase. 
	When {a high level} of social cohesion exists, cooperation always decreases as selection intensities increase. 
	With medium social cohesion (=0.5), if we increase the selection intensities from 0.05 to 0.5, cooperation decreases because it is not favored by evolution. However,  we would see {a} larger proportion of cooperation if selection intensity is 1 or 2. But again cooperation almost vanishes under too high selection intensities.}

Intuitively, when {the} selection intensity is too weak, then all the strategies almost make no difference in the evolution so we could not expect high levels of cooperation.
When selection intensity is too strong, the lucky defectors whose opponents are all cooperators would get the highest payoffs and gain more success.
So medium levels of selection intensity is best for cooperation.


\section{Conclusion}
\label{section-conclusion}

Cohesion is one of the most widely studied concepts in small-group performance an intra- and intergroup relations \citep{Evans1991group,Carron2000Cohesion,Chiocchio2009cohesion} as it is essential for teams where people of different talents and background meet together to collaborate on common tasks.
However, due to the complicated and elusive nature, researches have to face the difficulty in  defining, measuring, operationalizing, and experimentally manipulating cohesion \citep{Carron2000Cohesion}. 
And it always gives people the feeling of inconsistency when reviewing the literature relevant to cohesion\citep{Rosh2012Too}. 
Obviously, it is hard to unify all the previous definitions and measurements without giving new ones, which in turn only  {add} to the incoherence within this field of study. 
\citet{Friedkin2004Social} and \citet{Dion2000Group}  {call} for emphasis on the solid aspects of cohesion rather than trying to propose new ones, which is more practical and heuristic. 

One of the most significant and generally admitted  {findings} in the study of cohesion is the distinction between social and task cohesion, which is  referred as a milestone by \citet{Evans1991group}.
In general various meta-{analyses} of existing cohesion studies {are in agreement} that group performances are more strongly related with task cohesion than social cohesion \citep{Mullen1994The,Campbell2009Sticking,Chiocchio2009cohesion,Castano2013A,Grossman2015What}.
In these  {studies}, the definitions and measurements of cohesion are usually different among the literature,  {making} it hard to compare which factors are more crucial in the cohesion-performance relation.
\citet{Salas2015Human} conduct a meta-analysis and highlight that further study should give priority to social and task cohesion and incorporate dynamical and temporal factors into study.

Inspired by these  suggestions and  {empirical observations}, we compare in this work how social and task cohesion influence the emergence of cooperation  {by resorting to evolutionary game theory, since  in this} field, there  {has been significant success in understanding the dynamic processes} of group  {and agreement} formation and interactive strategies among group members \citep{Szabo2007evolutionary,Han:2014tl,HanJaamas2017,Newton2018Evolutionary} 
In particular, conditional dissociation mechanisms allow players to leave their groups once they are dissatisfied about their opponents and have been proven beneficial for cooperation to evolve \citep{Izquierdo2010,Qu2016Conditional,Qu2019How,Aktipis2004}.
We introduce group cohesion into this process to study how group cohesion  {influences} the performance of groups {and the evolutionary outcomes of cooperation}.

Unlike psychological study where the definitions of cohesion are largely dependent on the measurement the researchers choose, our definitions of social and task cohesion are provided according to the intrinsic nature of individual choices and decision making.
Social cohesion is defined as a strength that prevents players from leaving their groups regardless of the history of the play.
With higher levels of social cohesion, players are more tolerant towards defections and thus more refrain from leaving. 
Task cohesion is modeled as the likelihood that unsatisfied players would look back into the history before they choose to leave. 
If they were satisfied about the outcomes in the last round of play, they would choose to stay.

Our primary finding is that social cohesion has a negative effect {on} the group cooperation while task cohesion has a positive effect.
The difference of the effect of social and task cohesion could be illustrated from the perspective of positive assortativity, which is termed as the common characters of almost all mechanisms that {promote} cooperation, including conditional dissociation \citep{Eshel1331,doebeli2005models,Fletcher2009A,Izquierdo2010,Qu2019How}.
With either type of cohesion, dissatisfied cooperators are less likely to leave their groups, which means cohesion hinders the conditional dissociation mechanism.
However, task cohesion enhances positive assortativity by enforcing cooperative groups.
As cooperation is vulnerable to defections or mistakes, task cohesion helps players to distinguish if the defection is intentional. 
{A higher level} of task cohesion {enables} players to be more patient and thus protects cooperators from entering {a} more defective match pool.
So for organization{al} practitioners, when monitoring and improving cohesion, recalling the successful history regularly would benefit the team {for a} more productive team fulfillment.

We also discuss other parameters of the evolutionary dynamics and their influence on the cooperation, which is {a} standard {approach} in evolutionary game study. 
Our discussion is central on how it {is related to} other aspects of cohesion and its impact on group performances.
Compared with the complicated nature of cohesion, our model is rather simple  {yet} very powerful in that it is capable of revealing the underlying evolutionary dynamics related to cohesion,  {shedding light on the mechanisms based on which} cohesion influences group cooperation.

\section{Further discussion}
\label{further}
Cohesion is such an important and complex concept {that many aspects of it need to be further examined} in future.

The first one is to study other categories of games which enable us to explore cohesion from different angles.
Both our paper and \citet{Qu2019How} apply the public goods game, which is the most studied multi-person game.
But, apparently in reality, people engage in other {types} of interactions too.
Sometimes, people need to coordinate on a common action or idea, then the coordination game could be a better choice for study.
Other well-known game models including stag-hunt game, battle of sex, ultimatum game and etc. \citep{Szabo2007evolutionary} could also be applied to analyze different kinds of conflicts and interests among group members.

In our model, while  both social  and task cohesion are random and change over time, there is no direction of the changes.
So our model doesn't incorporate the reciprocal effect between cohesion and performances.
In previous  {works}, different or even contradictory patterns of cohesion are found in military units \citep{Bartone1999Cohesion,Siebold2006military}.
\citet{Grossman2015What} suggest that social cohesion emerges  {first then members shift attention} to task cohesion.
After they achieve more resources by accomplishing group goals, they in turn enhance social cohesion and finally both cohesion become stable over time.
So the second direction for further study would be to take the time inconsistency into consideration and in particular, the reciprocal effect between social and task cohesion.

Our models assume well-mixed population, and it is definitely necessary and interesting to investigate the influence of cohesion in structured populations or  {spatial networks where certain individuals are more likely to interact than} others.
{\citet{Aktipis2011Is} and \citet{Ichinose2018Network} study conditional dissociation mechanism on spatial networks and complex networks respectively.
There are multiple ways to introduce cohesion into their models.}
Cohesion may be referred  {to} as how often players want to leave, how far they can   {migrate to other parts in} the network, and when updating strategies, how far neighbors can be imitated.
Since different  {types of networks (e.g., homogeneous vs heterogeneous)} exhibit different tendencies in promoting cooperation, it would even more exciting to investigate how these topological properties interact with different aspects of cohesion.

Evolutionary game theory is a  powerful tool in analyzing group dynamics and behaviors.
Cohesion is an important concept in understanding team dynamics and performances.
{Our work suggests} it is promising to apply evolutionary game theory to study cohesion in more diverse directions, which may provide other interesting results and help us better understand how cohesion facilitates group performances.

\section*{Acknowledgements}
Xinglong Qu acknowledges support from the National Natural Science Foundation of China (NO. 71701058). 
 T.A.H. acknowledges  support from the Leverhulme Research Fellowship (RF-2020-603/9) and  Future of Life Institute (grant RFP2-154).

  \bibliographystyle{plainnat} 
\bibliography{cohesion}

\end{document}